%% file: akkus_et_al.tex
\documentclass[a4paper,12pt]{article} 

\usepackage{geometry,epsfig,subfigure}
\usepackage{amsmath}
\usepackage{amssymb}
\usepackage{graphics,epsfig,mydef}

\usepackage[square,sort&compress,comma,numbers]{natbib}

\usepackage[usenames]{color}
\usepackage{fancyhdr,url}   
\usepackage{setspace}  
\usepackage{multirow}   
\usepackage{rotating}    
\usepackage{colortbl}     
\usepackage{relsize} 
\usepackage{booktabs}
\usepackage{cancel}
\usepackage{enumerate}
\numberwithin{equation}{section}
\usepackage{soul} 
\usepackage{xcolor} 
\usepackage{pdflscape}
\usepackage{hyperref}
\usepackage[capitalise]{cleveref}
\usepackage{lineno}
\usepackage{graphicx}
\usepackage{etoolbox}
\usepackage{epstopdf}

\include{definitions}

\usepackage{amsthm}

\newcommand{\HRule}{\rule{0.9\linewidth}{0.2mm}}

\DeclareMathOperator\erf{erf}

\begin{document}
\renewcommand*{\thepage}{\arabic{page}}

\setstretch{1.3}

\begin{center}
\large
\textbf{Drifting mass accommodation coefficients: \textit{in situ} measurements from a steady state molecular dynamics setup\\}

\normalsize
\vspace{0.2cm}
Yigit Akkus$^{a}\symbolfootnote[1]{e-mail: \texttt{yakkus@aselsan.com.tr}}\!$, Akif Turker Gurer$^{a}$, Kishan Bellur$^{b,c}$\\
\smaller
\vspace{0.2cm}
$^a$ASELSAN Inc., 06200 Yenimahalle, Ankara, Turkey\\
$^b$Michigan Technological University, Houghton, MI 49931, USA\\
$^c$University of Michigan, Ann Arbor, MI 48109, USA\\
\vspace{0.2cm}
\end{center}

\begin{center} \noindent \HRule \\ \end{center}
\vspace{-0.6cm}
\begin{abstract}

\noindent 

A fundamental understanding of the evaporation/condensation phenomena is vital to many fields of science and engineering, yet there is much discrepancy in the usage of phase change models and associated coefficients. First, a brief review of kinetic theory of phase change is provided, and the mass accommodation coefficient (MAC, $\alpha$) and its inconsistent definitions are discussed. The discussion focuses on the departure from equilibrium; represented as a macroscopic ``drift" velocity. Then a continuous flow, phase change driven molecular dynamics setup is used to investigate steady state condensation at a flat liquid-vapor interface of argon at various phase change rates and temperatures to elucidate the effect of equilibrium departure. MAC is computed directly from the kinetic theory based Hertz-Knudsen (H-K) and Schrage (exact and approximate) expressions without the need for \emph{a priori} physical definitions, \emph{ad hoc} particle injection/removal or particle counting. MAC values determined from the approximate and exact Schrage expressions ($\alpha_{app}^{Schrage}$ and $\alpha_{exact}^{Schrage}$) are between 0.8 and 0.9, while MAC values from the H-K expression ($\alpha^{H-K}$) are above unity for all cases tested. $\alpha_{exact}^{Schrage}$ yields values closest to the results from transition state theory [\textit{J Chem Phys}, \textbf{118}, 1392--1399 (2003)]. The departure from equilibrium does not affect the value of $\alpha_{exact}^{Schrage}$ but causes $\alpha^{H-K}$ to vary drastically emphasizing the importance of a drift velocity correction. Additionally, equilibrium departure causes a non-uniform distribution in vapor properties. At the condensing interface, a local rise in vapor temperature and a drop in vapor density are observed when compared with the corresponding bulk values. When the deviation from bulk values are taken into account, all values of MAC including $\alpha_{exact}^{Schrage}$ show a small yet noticeable difference that is both temperature and phase change rate dependent.

\vspace{0.2cm}


\end{abstract}
\vspace{-0.6cm}
\begin{center} \noindent \HRule \\ \end{center}

\pagebreak

\section{Introduction}
\label{sec:intro}

Classical kinetic theory is a statistical description of the behavior of gases based on velocities of the constituent molecules and has provided the basis for modeling liquid-vapor phase change. Under equilibrium conditions, the vapor in the vicinity of the liquid-vapor interface can be approximated as a perfect (ideal) gas and the velocity distribution of the vapor molecules follows a Maxwell Boltzmann distribution \cite{chapman1990,carey2018}. This velocity distribution leads to an expression for the \emph{maximum} collision frequency with a planar surface. Phase change is a dynamic process and a pure liquid-vapor system undergoes simultaneous condensation and evaporation. A net phase change flux is generally expressed as an algebraic sum of evaporation and condensation fluxes. 

Mass accommodation coefficients (MAC) were introduced to account for deviation from the kinetic theory predicted \emph{maximum} flux. The deviation is attributed to reflection of vapor molecules at the interface. There has been much discrepancy in both definition and reported values of MAC \cite{kryukov2011,marek2001,persad2016}. For example, Marek and Straub \cite{marek2001} reported a spread of 3 orders of magnitude in prior published values for water and Kryukov and Levashov \cite{kryukov2011} reported 3 different definitions of MAC. This is further complicated by the fact that additional modifications have been made to the original equations. Coefficients values reported from a particular kinetic theory expression cannot be used interchangeably with a different expression due to the lack of a standard definition for the coefficient. This reduces the coefficient to an empirical fitting parameter that is not universally applicable. Incognizance of fundamental assumptions is a possible reason for much controversy regarding both the applicability of kinetic theory expressions and the corresponding coefficients \cite{persad2016}. In the rest of this section, we review the most common kinetic theory expressions from a fundamental standpoint and explore the prior published values of the accommodation coefficient.

\subsection{Kinetic model for liquid-vapor phase change}
\label{sec:kinetic}

\begin{figure} [h]
\includegraphics[width=2.5in]{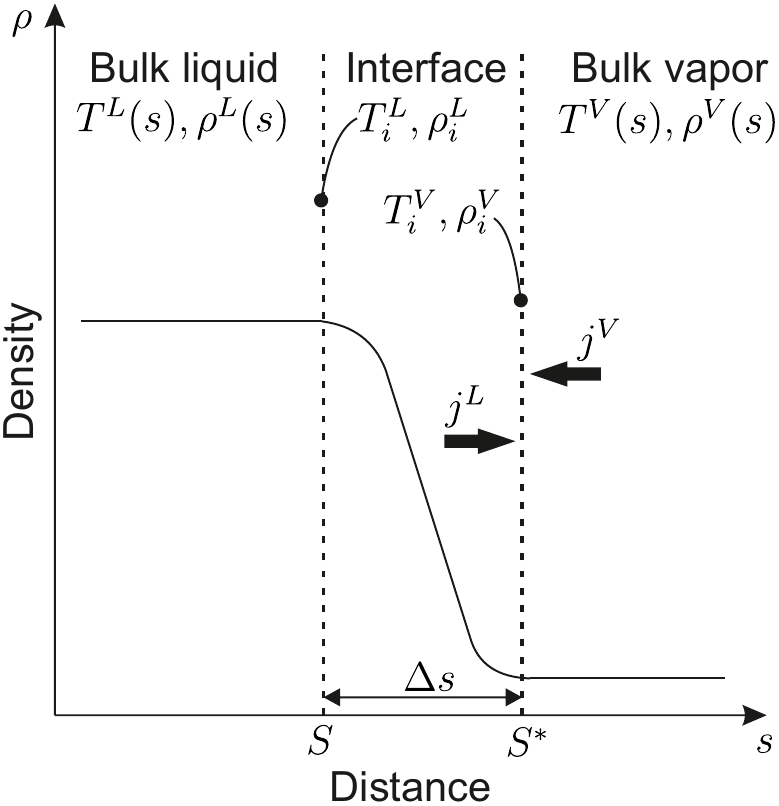}
\centering
\caption{\label{fig:fig_int_noneq} Thermodynamic quantities used for the diffuse interface during net evaporation or condensation process. $S$ is the intersection of the bulk liquid and the interfacial region and $S^*$ is the intersection of the bulk vapor and the interfacial region.}
\end{figure}

Assuming the distribution of vapor molecules to be Maxwellian and treating the vapor as an ideal gas, mass flux crossing a hypothetical plane surface can be expressed as \cite{hertz1882}:

\begin{equation} \label{eq:coll_f}
 \ j^{V} = \rho^{V} \sqrt{\frac{k_b T^{V}} {2 \pi m }}
\end{equation} 

\noindent where $\rho^{V}$, $k_b$, $m$, and $T^{V}$ are vapor density, Boltzmann constant, mass of a molecule, and vapor temperature, respectively, and the superscript $V$ denotes the vapor phase. \begin{figure} [h]
\includegraphics[width=2.5in]{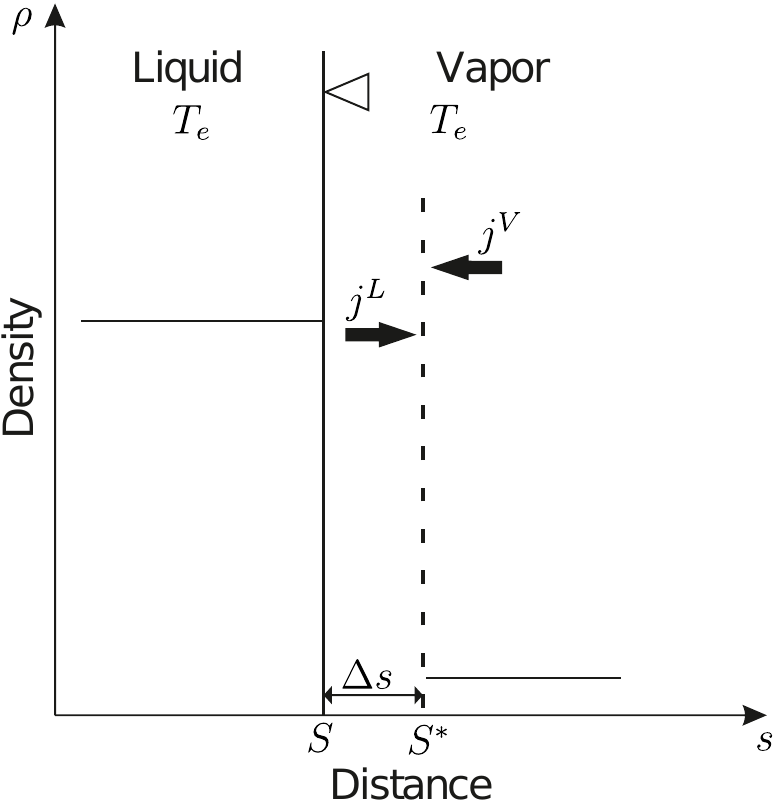}
\centering
\caption{\label{fig:fig_int_eqm} Liquid-vapor interface at equilibrium. Solid vertical line demonstrates the interface defined based on the sharp interface approach. Vertical dash line is a hypothetical surface close to the sharp interface. $\Delta$s is assumed to be infinitesimally small.}
\end{figure} In order to extrapolate this expression to interphase mass transport such as a liquid-vapor system, we must first define the liquid-vapor interface. In many prior studies, the interface was assumed to be sharp \cite{schrage1953}. Since Eq.~\ref{eq:coll_f} is theoretically only applicable in the vapor phase, a hypothetical surface ($S^*$) close to the sharp interface ($S$) on the vapor side is considered (Fig.~\ref{fig:fig_int_eqm}) \cite{carey2018}. The condensation flux $j^{V}$ is the number of vapor molecules crossing the surface $S^*$ in the negative $s$-direction and the evaporation flux $j^{L}$ is the number of vapor molecules crossing the surface $S^*$ in the positive $s$-direction as shown in Fig.~\ref{fig:fig_int_eqm}. $S^*$ is generally assumed to be infinitesimally close to $S$ but a formal description is lacking \cite{carey2018}. In reality, the interface is not sharp but diffuse and an ``interfacial region" with a density gradient exists \cite{standart1958,yasuoka1994argon,matsumoto1994methanol,nagayama2003,meland2004noneq,kryukov2011}. In the interfacial region, both $\rho^V$ and/or $T^V$ would vary with $s$ and hence, a formal definition of $S$ and $S^*$ is warranted. In order to aid in both accuracy and consistency, $S$ could be defined as the intersection of the interfacial region and the bulk liquid and $S^*$ could be defined as the intersection of the interfacial region and the bulk vapor (Fig.~\ref{fig:fig_int_noneq}). 

In equilibrium, the evaporation flux is equal to the condensation flux, $j^{V}$ = $j^{L}$. There is no temperature jump across the interface and the liquid-vapor system is saturated. The vapor density in equilibrium is equal to the vapor saturation density at the corresponding saturation temperature. Hence,

\begin{equation} \label{eq:equilibrium_flux}
 j^{V} = j^{L} = \rho^V_{sat}(T_e) \sqrt{\frac{k_b T_e} {2 \pi m}}
\end{equation} 

\noindent where superscript $L$ and subscript $e$ denotes the liquid phase and equilibrium, respectively.

During net evaporation or condensation, $j^V$ and $j^L$ are not equal. We entertain the possibility that the vapor temperature could be different from the liquid temperature and undergo a transition in the interfacial region similar to density (Fig.~\ref{fig:fig_int_noneq}). Additionally, interfacial temperature and densities could be different from the corresponding bulk phase values. The interfacial properties are shown using a subscript $i$. A net phase change rate can be defined as the difference between the evaporation and condensation fluxes. The condensation flux ($j^V$) can be expressed by Eq.~\eqref{eq:coll_f}. The estimation of the evaporation flux is not straightforward. A common approach is to consider an absolute rate of evaporation based an liquid-vapor system equilibrated at the liquid interfacial temperature ($T^L_i$) \cite{schrage1953}. The absolute rate of evaporation is then given by Eq.~\eqref{eq:equilibrium_flux} where $T_e$ is replaced by $T^L_i$. Although there are several arguments to support the equilibrium approach to estimate evaporation flux \cite{schrage1953,carey2018}, several authors have argued against it \cite{algie1978} making this a point of open debate.

Assuming the variations of temperature and density are negligible in the vapor phase (i.e. $T_i^V=T^V$ and $\rho_i^V=\rho^V$), net phase change flux is an algebraic sum of evaporation and condensation flux and is given by Hertz relation \cite{hertz1882}:

\begin{equation} \label{eq:hertz}
\dot{m}^{''}=\sqrt{\frac{k_b}{2 \pi m}}\left(\rho^V_{sat}(T^L_i)\sqrt{T_i^L}-\rho^V\sqrt{T^V}\right)
\end{equation}

\noindent where $\dot{m}^{''}$ is the net mass flux. In the Hertz formulation (Eq.~\eqref{eq:hertz}), condensation flux is dependent on the local thermodynamic quantities (density and temperature) on the vapor side, while the evaporation flux is dependent on the same quantities but on the liquid side. In other words, the rates of the concurrent condensation and evaporation processes only depend on the properties of their respective phases. A common argument made to support this approximation is that $\Delta s$ is infinitesimally small. If $\Delta s$ is smaller than 1 mean free path then any molecule that evaporates from the bulk liquid must pass through $S^*$ before interacting with a vapor molecule thereby preserving the liquid properties as it passes through $S^*$. The Hertz approach has been criticized, since it actively decouples any interaction between the two fluxes \cite{persad2016}. In essence, the Hertz equation provides the theoretical maximum phase change flux possible, since molecular reflection at $S^*$ was not incorporated.

Early experiments \cite{eames1997,marek2001} consistently measured phase change rates lower than that predicted by the Hertz equation. This is generally attributed to reflection of vapor molecules at the interface. When a vapor molecule is incident on the interface, it can interact in three ways: (i) the molecule can condense (i.e., the vapor molecule is absorbed into the bulk of the liquid), (ii) the molecule can be reflected back into the vapor space or (iii) the molecule can displace a liquid molecule thereby undergoing a simultaneous condensation-evaporation process. Vapor reflection reduces both the condensation and evaporation mass flux. In order to account for this deviation from the theoretical maximum (Eq.~\eqref{eq:hertz}), evaporation and condensation coefficients were introduced \cite{knudsen1915} and the result is widely known as the Hertz-Knudsen equation: 

\begin{equation} \label{eq:HK}
\dot{m}^{''}=\sqrt{\frac{k_b}{2 \pi m}}\left(\alpha_e\rho^V_{sat}(T^L_i)\sqrt{T_i^L}-\alpha_c\rho^V\sqrt{T^V}\right)
\end{equation}

\noindent where $\alpha_e$ is the evaporation coefficient and $\alpha_c$ is the condensation coefficient.

\subsubsection*{\underline{Coefficient definition(s)}}
Before any discussion of the coefficient values, we must first develop a definition. There is much inconsistency in the definition of the coefficients reported by prior studies. Most prior definitions could be grouped into 5 major categories: 
\begin{description}
\item [Definition 1] Ratio of measured rate to calculated rate \cite{xia1994,yasuoka1994argon,matsumoto1994methanol,eames1997,matsumoto1998,anisimov1999,holyst2015}.
\item [Definition 2] Probability of capture or absorption \cite{barrett1992,tsuruta2005,cheng2011}.
\item [Definition 3] Ratio of condensed molecules to incident molecules \cite{yasuoka1995,tsuruta1999,rosjorde2001,nagayama2003,tsuruta2004,meland2004noneq,liang2017}.
\item [Definition 4] Correction factor for quality of phase boundary \cite{standart1958}.
\item [Definition 5] Efficiency of molecules adhering to or abandoning the surface \cite{fuster2010}.
\end{description}

\textbf{Definition 1} inherently makes the coefficient dependent on kinetic theory and is not tied to a physical description. The measured rate is compared to the kinetic theory predicted rate and the coefficient is determined by comparison. This is convenient when using experiments to determine the coefficient. However, when molecular dynamics or other purely computational methods are used, other definitions are generally utilized. \textbf{Definitions 2--5} are independent of kinetic theory of phase change. The biggest open debate is whether the coefficient is an intrinsic property of the liquid-vapor system or just a fudge factor to kinetic theory expressions. 

Let us consider the case where \textbf{Definition 1} is used to calculate the coefficient. The primary complication in evaluating Eq.~\eqref{eq:HK} is that the interfacial liquid temperature ($T_i^L$) and both kinetic coefficients are unknown. Even if $T_i^L$ is measured or approximated, there remain two unknowns, $\alpha_c$ and $\alpha_e$, in the expression for $\dot{m}^{''}$. For sake of closure it is common practice to assume that the condensation coefficient is equal to the evaporation coefficient ($\alpha_c = \alpha_e = \alpha$) \cite{wayner1976,barnes1986,wayner1991,wayner1999,marek2001,plawsky2008,holyst2013nanoscale,liang2017}. Under this assumption the only remaining coefficient in Eq.~\eqref{eq:app_HK} is $\alpha$, which is referred to as the \emph{mass accommodation coefficient (MAC)}.

\begin{equation} \label{eq:app_HK}
\dot{m}^{''}=\alpha\sqrt{\frac{k_b}{2 \pi m}}\left(\rho^V_{sat}(T^L_i)\sqrt{T_i^L}-\rho^V\sqrt{T^V}\right)
\end{equation}

Another popular form of the Hertz-Knudsen equation is
\begin{equation} \label{eq:app_HK_pressure}
\dot{m}^{''}=   \alpha \frac{1}{\sqrt{2 \pi R}}\left(\frac{p_{sat}|_{T_i^L}}{\sqrt{T_i^L}}-\frac{p^V}{\sqrt{T^V}}\right)
\end{equation}

\noindent where $p$ is pressure and $R$ is the specific gas constant. This is equivalent to Eq.~\eqref{eq:app_HK} when an ideal gas approximation is made to convert densities to pressures. This poses a concern, since phase change is an inherent non-equilibrium process. The velocity distribution has the potential to deviate from the equilibrium Maxwellian which makes the applicability of the ideal gas expression to the vapor close to the interface suspect. In the rest of the study, we refer to Eq.~\eqref{eq:app_HK}, as the Hertz-Knudsen (H-K) equation recognizing that two assumptions are made: (i) equality of phase change coefficients; $\alpha_c = \alpha_e = \alpha$ and (ii) uniformity in vapor properties; $T_i^V=T^V$ \& $\rho_i^V=\rho^V$. The density form of the kinetic theory expression is retained so that modifications to account for potential departure from equilibrium could be introduced.

\subsubsection*{\underline{Departure from equilibrium}}
Under equilibrium conditions, the evaporation flux is equal and opposite to the condensation flux (i.e. the net flux is zero) and the velocity distribution is a perfect Maxwellian. Phase change is an inherently non-equilibrium process and Schrage \cite{schrage1953} argued that during steady phase change there is a net macroscopic velocity of the vapor molecules either towards or away from the interface. This is also referred to as a ``drift'' velocity. Drift velocity was superimposed with the Maxwell Boltzmann distribution to develop a correction factor ($\Gamma$). Schrage's formulation can be expressed as,

\begin{subequations} \label{eq:exact_Schrage_all}

\begin{equation} \label{eq:exact_Schrage}
\dot{m}^{''}=  \alpha \sqrt{\frac{m}{2 \pi k_B}}\left(\frac{\rho_{sat}|_{T_i^L}}{\sqrt{T_i^L}}-\Gamma (a)\frac{\rho^V}{\sqrt{T^V}}\right)
\end{equation}

\noindent where $a$ is the ratio of the drift velocity ($w_0$) to the mean thermal velocity of the vapor molecules (Eq.~\eqref{eq:defn_a}) and $\Gamma$ is the correction factor (Eq.~\eqref{eq:defn_Gamma}):

\begin{equation} \label{eq:defn_a}
a=\frac{w_0}{\sqrt{2k_B T^V/m}} 
\end{equation}

\begin{equation} \label{eq:defn_Gamma}
\Gamma(a)=\exp(-a^2)-a\sqrt{\pi}[1-\erf{(a)}]
\end{equation}

\end{subequations}

\noindent where $w_0$ is the drift velocity in Eq.~\eqref{eq:defn_a} and is given by $w_0=\dot{m}^{''}/\rho^V$, where $\rho^V$ is the vapor density. If the drift velocity is small in comparison to the thermal velocity, Eq.~\eqref{eq:defn_Gamma} reduces to $\Gamma(a)\approx 1+a\sqrt{\pi}$ \cite{carey2018}. If the ideal gas expression is used to evaluate $\rho^V$, then, in the limit of small $a$, the original Schrage expression (Eq.~\eqref{eq:exact_Schrage}) can be reduced to Eq.~\eqref{eq:app_Schrage} \cite{barrett1992,carey2018,liang2017}.

\begin{equation} \label{eq:app_Schrage}
\dot{m}^{''}=  \frac{2\alpha}{2-\alpha} \sqrt{\frac{m}{2 \pi k_B}}\left(\frac{\rho_{sat}|_{T_i^L}}{\sqrt{T_i^L}}-\frac{\rho^V}{\sqrt{T^V}}\right)
\end{equation} 

While the kinetic factor in H-K equation (Eq.~\eqref{eq:app_HK}) is simply $\alpha$, it is $2\alpha/(2-\alpha)$ in the approximate Schrage expression \footnote{A majority of papers erroneously refer to Eq.~\eqref{eq:app_Schrage} as the original Schrage expression. Although it is derived from the original equation developed by Schrage \cite{schrage1953}, there are two inherent assumptions: (i) the drift velocity of the vapor molecules is small in comparison to the mean thermal velocity, and, (ii) ideal gas equation is used to evaluate vapor density.} (Eq.~\eqref{eq:app_Schrage}). A few researchers assume $\alpha$=1 in Eq.~\eqref{eq:app_HK} while others assume the same but in Eq.~\eqref{eq:app_Schrage}  \cite{do2008,du2011,akkus2017,alipour2019,akdag2020}. In such a case, the approximate Schrage expression (Eq.~\eqref{eq:app_Schrage}) would predict twice the mass flux predicted by H-K equation (Eq.~\eqref{eq:app_HK}) for the same value of $\alpha$ = 1. A value of unity is just a theoretical upper limit (i.e., it is the limit predicted by Hertz relation, Eq.~\eqref{eq:hertz}) and most prior studies have reported values less than unity \cite{marek2001}.

During non-equilibrium phase change, Fang and Ward \cite{fang1999} were first to report experimental evidence of a temperature discontinuity. Ward and Stanga \cite{ward2001} later reported the gas temperature rise near a condensing interface (i.e. $T^V>T_i^V$). This suggests that equilibrium departure has the potential to cause a drift between the interfacial vapor properties and the corresponding bulk values, i.e $T_i^V$ and $\rho_i^V$ may be different from $T^V$ and $\rho^V$, respectively. The commonly made assumption of uniform vapor phase properties is invalid at high phase change rates and has the potential to lead to erroneous values of MAC.

\subsection{Prior measurements of MAC}
\label{sec:MD}

Experimentally determined values of MAC have been highly inconsistent \cite{barnes1986,marek2001,davis2006}. For water alone, the reported values vary by almost three orders of magnitude \cite{marek2001}. To determine MAC from experiments, $\dot{m}^{''}$ and $T_{i}^L$ must be measured with a high degree of accuracy and this poses several experimental challenges; the first of which is  the existence of large temperature jumps at the interface \cite{persad2016}.  Second, if the interface is not perfectly flat, additional factors could alter both $\dot{m}^{''}$ and $T_{i}^L$ considerably \cite{bellur2020}. Lastly, the presence of impurities further alters the shape of the interface and thereby the local properties. The experimental discrepancy in prior measured values of MAC have been attributed to difficulty in measuring interfacial temperature, dynamic surface tension, renewing/re-wetting surfaces, and trace impurities in the liquid \cite{davis2006,marek2001,barnes1986}.
 
Molecular dynamics (MD) simulations are an alternative to mitigate the aforementioned experimental challenges and have been widely used to investigate the phase change phenomenon. Consequently, a massive body of literature exists for the prediction of MAC by MD simulations. In the proceeding paragraphs, we highlight several influential studies.   

Adopting \textbf{Definition 1} for the calculation of MAC, Yasuoka, Matsumoto, and Kataoka \cite{yasuoka1994argon,matsumoto1994methanol} estimated MAC values for the condensation of argon and methanol as $\sim$0.8 and $\sim$0.2, respectively, in their equilibrium simulations. Later, Matsumoto \cite{matsumoto1998} reported MAC values for non-equilibrium simulations and pointed out the inverse relation between MAC values and the temperature. This dependence was also reported in many subsequent studies \cite{matsumoto1998,anisimov1999,nagayama2003,ishiyama2004,tsuruta2004,tsuruta2005,holyst2009,cheng2011,nagayama2015,liang2017}.  

Tsurata \textit{et al.} \cite{tsuruta1995,tsuruta1999} investigated condensation probability of argon atoms, which were injected to the system and targeted to a condensing surface, and came up with a velocity dependent coefficient formulation ($\sigma_c$) for the condensation of individual argon atoms. Then using transition state theory \cite{nagayama2003,tsuruta2005}, they reported a general expression for the average condensation coefficient of all atoms ($\overline{\sigma_c}$) expressed by the specific volume ratio between liquid and vapor (translational length ratio). Our results are compared to the values found by using their formulation later in Section~\ref{sec:non-eq}. 

Cheng \textit{et al.} \cite{cheng2011} conducted MD simulations to show the effects of molecular composition on the evaporation. They reported values for a single coefficient, namely condensation coefficient on an evaporating interface, and demonstrated that dimers and trimers had higher coefficients than monomers. However, coefficients were demonstrated to collapse onto a master curve when plotted against a translational length ratio, a concept suggested in \cite{nagayama2003}.

Meland \textit{et al.} \cite{meland2004noneq} reported evaporation ($\alpha_e$) and condensation ($\alpha_c$) coefficients of LJ-spline fluid on both condensing and evaporating interfaces. They showed that these coefficients are not equal outside the equilibrium and depends on the drift velocity. Nagayama \textit{et al.} \cite{nagayama2015} investigated the condensation/evaporation coefficients of some straight-chain alkanes (butane, dodecane, and octane) and estimated values consistent with the transition state theory.

Two studies of Ho{\l}yst \textit{et al.} \cite{holyst2013nanoscale,holyst2015} attracted the attention of the community by reporting MAC values higher than unity. Authors adopted \textbf{Definition 1} for the calculation of MAC value and used Eq.~\eqref{eq:app_HK_pressure} to estimate the theoretical prediction of the phase change rate. However, they assumed thermal equilibrium at the interface ($T_i^L = T^V$). In 2016, Persad and Ward \cite{persad2016} published an extensive review on the evaporation and condensation coefficients and discussed commonly used simplifying assumptions during the estimation of these coefficients, among which the thermally equilibrated interface was criticized since the temperature discontinuity has been already revealed in many experimental and numerical studies. Persad and Ward \cite{persad2016} also introduced relations for evaporation and condensation coefficients based on statistical rate theory of quantum mechanics and demonstrated that coefficient values are not bounded by unity. They concluded that H-K relation is incomplete due to the decoupling of the two interacting phases.

A number of prior molecular dynamics studies calculated the coefficient by tracking particles that cross a hypothetical plane and determining the rate of reflection \cite{yasuoka1994argon,matsumoto1994methanol,tsuruta1995,tsuruta1999,nagayama2003,tsuruta2005,cheng2011,liang2017}. This method raises concerns over the appropriateness of the time period for which the particle is tracked and the location of the hypothetical plane with respect to the interfacial region. Cheng \textit{et al.} \cite{cheng2011} referred to picking the time period as an \emph{ad-hoc} approach. Calculation of MAC using molecular dynamics in the past has been primarily through use of equilibrium simulations \cite{matsumoto1992,yasuoka1994argon,matsumoto1994methanol,tsuruta1995,matsumoto1998,tsuruta1999,nagayama2003,nagayama2015}. As discussed earlier, an equilibrium approach results in a net zero phase change flux and a net zero drift velocity. Hence, using an equilibrium molecular dynamics approach does not include the effect of drift velocity on phase change. Under non-equilibrium conditions the MAC determined by particle tracking could be different from those at equilibrium \cite{meland2004noneq} and a velocity correction in the vapor is necessary \cite{liang2017}. Lastly, while a considerable number of non-equilibrium MD setups in previous studies was inherently transient \cite{yu2012,holyst2009,ishiyama2004,tsuruta2005,cheng2011,liang2017,liang2018}, the ones with the steady state phase change process \cite{hafskjold1995,meland2004noneq,holyst2013nanoscale,holyst2015} used \textit{ad hoc} methods such as removing and/or injecting particles at prescribed regions of the simulation domain to sustain the continuous phase change process.

\subsection{Current study}
\label{sec:study}

Extreme caution must be used when using coefficients estimated  through equilibrium simulations or particle tracking methods in kinetic theory expressions for three reasons: i) there is huge ambiguity in the definition of the coefficients and prior published values lack universal applicability to all kinetic theory based expressions, ii) departure from equilibrium and the effect of drift velocity must be accommodated by both the simulation technique and the kinetic theory expression, and iii) the kinetic theory approach to phase change is built on multiple assumptions (that involves decoupling of two actively interacting phases \cite{persad2016}), and the simulation may not take this into account. The above problems can be avoided if the coefficients are back-calculated from H-K or Schrage's relation using data from a non-equilibrium, steady state molecular dynamics simulation. Since these equations are used in many engineering applications, obtaining coefficients that have direct applicability in these kinetic theory expressions is critical.

We aim to present the first NEMD study calculating the coefficients in a steady state manner without the need for \emph{ad hoc} particle injection/removal methods. Instead of assuming a definition for MAC and attributing it as a function of the drift velocity, we compute the absolute value of MAC by direct comparison of data from a non-equilibrium, steady state MD simulation with kinetic theory models of phase change (H-K and Schrage). Non-equilibrium phase change is shown to result in both drift velocity and a non-uniformity in vapor-phase properties. Based on the results from our simulations we discuss the validity of the different definitions for MAC and outline the effect of equilibrium departure on coefficients derived from both H-K and Schrage expressions.

\section{Methodology}
\label{sec:methodology}

\begin{figure} [h!]
\noindent
\makebox[\textwidth]{\includegraphics[width=6in]{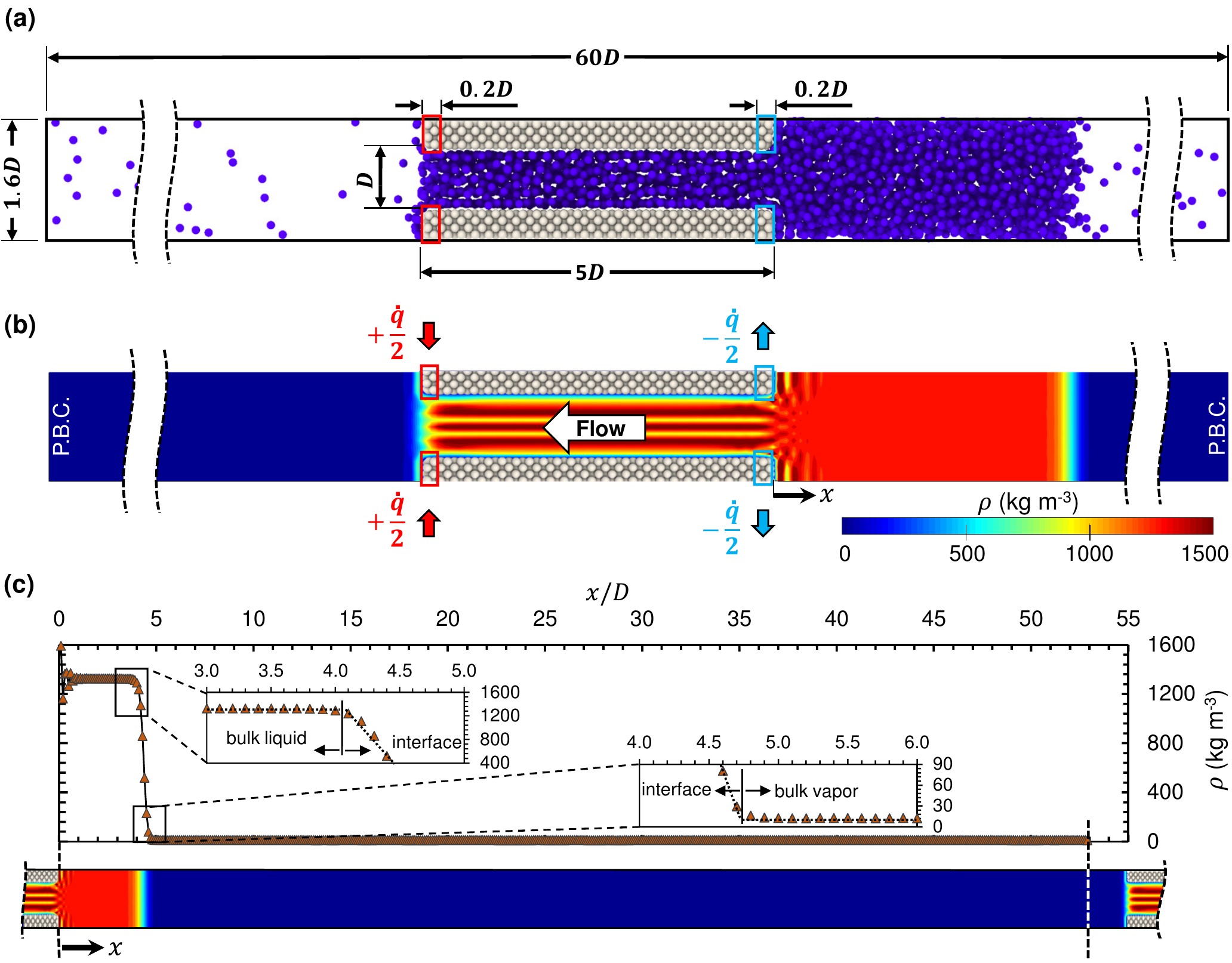}}
\caption{\label{fig:domain} (a) A snapshot of the computational setup at $90 \unit{K}$. Distance between the channel walls ($D$) is $1.96 \unit{nm}$ and the other dimensions in the simulation domain are proportional as shown in the figure. Depth of the simulation domain is $3.72 \unit{nm}$. The walls are composed of four atom layers and have a thickness of $0.59 \unit{nm}$. (b) 2-dimensional density distribution of the fluid (argon) after sufficient time averaging of MD results. Molecular layering, an experimentally observed phenomenon \cite{heslot1989,cheng2001} resulting from wall-force-field effect is apparent in the liquid phase between the walls. Red and blue rectangles enclose the solid atoms, which are subjected to the equal energy injection and extraction processes, respectively. (c) 1-dimensional density distribution of the fluid in longitudinal direction. Vertical bins with a thickness of $0.1D$ are used to calculate the average density along $x$-direction, which starts at the side edges of the walls near the condenser side. Periodicity at the side boundaries enables a continuous gas phase along the $x$-direction. Data collection is ceased at a distance of $2D$ from the evaporating interface to eliminate the possibility of any density variation in the vapor phase. Insets show the close-up view of the density distribution near the intersections of interfacial region with the bulk liquid and vapor phases.}
\end{figure}

Phase change driven nanopump technique, an MD simulation proposed by \cite{akkus2019molecular} and utilized in different applications \cite{akkus2019atomic,akkus2019first,akkus2019modeling}, is used to investigate the steady state condensation process at a flat liquid-vapor interface. The computational setup consists of two parallel walls composed of Platinum (Pt) atoms (gray spheres in Fig.~\ref{fig:domain}). In the transverse direction, size of the simulation domain is determined by the outermost Pt layers of each wall. In the longitudinal direction, simulation domain extends substantially beyond the walls. When Argon (Ar) atoms are introduced into the system (blue spheres in Fig.~\ref{fig:domain}a), they preferentially condense in the space between the two walls and form a liquid bridge due to the attraction of Ar atoms with the wall atoms. Ar atoms are placed asymmetrically in the computational domain such that when condensed into liquid phase, one of the free surfaces are pinned between the edges of the walls by means of capillarity, while the other free surface forms away from the wall edges at the opposite side, thereby creating a liquid slab as shown in Fig.~\ref{fig:domain}a. The number of Ar atoms to be introduced is selected such that the thickness of the liquid slab attached to the walls at one end is appreciably higher than $2.5 \unit{nm}$, which was reported as a limit for the presence of the effects of disjoining pressure \cite{yu2012}. The rest of the simulation domain is occupied by the vapor phase of Ar. Periodic boundary conditions are applied in all directions. Periodicity in transverse direction renders the system analogous to a liquid block placed on a semi-infinite wall within a sufficiently large vapor medium. Moreover, sufficient thickness of the liquid slab prevents the formation of a curved interface at the free surface. Consequently, a flat interface is achieved at the surface of the liquid slab. This allows for direct applicability of kinetic models of phase change at planar interfaces.  

In what follows, simulation steps are described in detail. Then the method used in the detection of interfacial region is explained. Moreover, further details of MD simulations are provided in Supplementary Material. 

\subsection{Simulation steps}
\label{sec:steps}

\subsubsection{Thermostat application period}
\label{sec:NVT}

To stabilize the system at the prescribed temperature, constant NVT ensemble (constant atom number, volume, and temperature) is applied to all atoms by the Nos\'e-Hoover thermostat method for $60 \unit{ns}$ except the atoms in outermost layers of the walls, which are always fixed at their lattice positions to preserve the shape of the system. 

\subsubsection{Equilibration period}
\label{sec:NVE}

Following the thermostat application, microcanonical (NVE) ensemble (constant atom number, volume, and energy) is applied to Ar atoms for $120 \unit{ns}$ to equilibrate the system. During this equilibration period, wall atoms are still subjected to the thermostat. At the end of this stage, thermally equilibrated and statistically stable liquid-vapor mixture is achieved. With time averaging, liquid-vapor interfaces become apparent in the system as shown in Fig.~\ref{fig:domain}b. To determine the thermodynamic properties of the saturated fluid in the present computational setup, data are collected from the equilibrated system during this stage. 

\subsubsection{Heating and cooling period}
\label{sec:NEMD}

At the liquid-vapor interfaces of the equilibrated system, simultaneous evaporation and condensation of fluid atoms take place; however, the net rate of phase change is always zero. In order to create an interface with a net phase change rate, an energy exchange mechanism should be established. This mechanism is created by equally heating and cooling the solid atoms at the opposite ends of the nanochannel shown in Fig.~\ref{fig:domain} by red and blue rectangles, respectively, while Ar atoms are subjected to microcanonical (NVE) ensemble. Cooled atoms in the liquid slab lead to the net condensation while at the other end, the pinned interface is heated leading to a net evaporation. Condensed atoms are transported from the condensing interface to the evaporating interface \textit{via} both Laplace pressure difference and molecular/atomic diffusion through the nanochannel \cite{akkus2019molecular}. In the absence of a meniscus, molecular/atomic diffusion near the walls (associated with the solid-liquid surface tension gradient \cite{akkus2019atomic,yesudasan2016}) is responsible for the pumping of the liquid. Periodic boundary conditions enable the steady and continuous flow of Ar not only in liquid phase but also in vapor phase. Therefore, the system behaves like a nanopump, continuously pumping the fluid in a prescribed direction with zero overall heat transfer due to equal heating and cooling. Utilization of the nanopump technique to create a statistically stable, flat interface with steady interfacial mass transfer makes the current computational setup unprecedented. This truly steady state setup does not utilize any \textit{ad hoc} treatment to sustain the steady operation and enable collecting data without any time restriction. In the present study, data are collected for $600 \unit{ns}$ during the phase change. However, this duration can be further increased to reduce the uncertainty associated with the data. 
 
\subsection{Interface detection}
\label{sec:int_detect}
In the absence of the wall-force-field effect (i.e. away from the walls), density distribution of the fluid along the transverse direction is homogeneous. The liquid-vapor interface is intentionally positioned away from the walls by adjusting the number of atoms used in the simulations. Both the interfacial region and the bulk vapor phase are free from any density variation in the transverse direction. Therefore, vertical bins are used to calculate the 1-dimensional distribution of density (see Fig.~\ref{fig:domain}c) along the longitudinal direction. Liquid-vapor interfacial region is characterized by a density gradient. A piecewise fit to the density profile is made considering the bulk phases and the center of the interfacial region. The interfacial region is delineated by intersections of the extrapolated piecewise fits similar to the approach in \cite{meland2004noneq}. These denote points of inflection (i. e. change in slope) of the density profile. (see the insets in Fig.~\ref{fig:domain}c). 

\subsection{Calculation of MAC values}
\label{mac}
Mass accommodation coefficients are calculated directly from the kinetic models of phase change (see Section~\ref{sec:kinetic}): i) H-K equation (Eq.~\eqref{eq:app_HK}), ii) exact Schrage relation (Eq.~\eqref{eq:exact_Schrage_all}), and iii) approximate Schrage relation (Eq.~\eqref{eq:app_Schrage}). As stated earlier, these equations are generally expressed in terms of pressure and temperature. However, the pressure, especially in the liquid phase, is subjected to considerable fluctuations \cite{copley1975,rosjorde2001}, which negatively affects the estimation of pressure. On the other hand, estimation of density is more reliable and straightforward. Further, the original kinetic theory formulation was based on density and we preserve the original formulation here without an ideal gas assumption. It should be noted that these equations (Eqs.~\eqref{eq:app_HK}--\eqref{eq:app_Schrage}) were developed for a net evaporation rate. We consider only the flat condensing interface in this study. Hence, the equations are re-arranged for net condensation: 

\begin{equation} \label{eq:HK_cond}
\dot{m}^{''} =\alpha^{H-K}   \sqrt{\frac{k_B}{2 \pi m}}  \left(\rho^V \sqrt{T^V} - \rho_{sat}|_{T_i^L} \sqrt{T_i^L}\right)
\end{equation}

\begin{equation} \label{eq:exact_Schrage_cond}
\dot{m}^{''} =\alpha_{exact}^{Schrage}   \sqrt{\frac{k_B}{2 \pi m}}  \left(\Gamma (-a) \rho^V \sqrt{T^V} - \rho_{sat}|_{T_i^L} \sqrt{T_i^L}\right)
\end{equation}

\begin{equation} \label{eq:app_Schrage_cond}
\dot{m}^{''} =\frac{2 \alpha_{app}^{Schrage}}{2 - \alpha_{app}^{Schrage}}    \sqrt{\frac{k_B}{2 \pi m}}\left(\rho^V \sqrt{T^V} - \rho_{sat}|_{T_i^L} \sqrt{T_i^L}\right)
\end{equation}

\noindent where mass accommodation coefficients (MACs) are designated according to the equations, in which they are used: i) $\alpha^{H-K}$ is the MAC for Hertz-Knudsen equation, ii) $\alpha_{exact}^{Schrage}$ is the MAC for exact Schrage relation, and iii) $\alpha_{app}^{Schrage}$ is the MAC for approximate Schrage relation. These MACs can be explicitly calculated based on their respective equations as follows:

\begin{equation} \label{eq:alpha_HK}
\alpha^{H-K}=\dot{m}^{''}   \sqrt{\frac{2 \pi m}{k_B}}{\left(\rho^V \sqrt{T^V} - \rho_{sat}|_{T_i^L} \sqrt{T_i^L}\right)}^{-1}
\end{equation}

\begin{equation} \label{eq:alpha_exact_Schrage}
\alpha_{exact}^{Schrage}=\dot{m}^{''}   \sqrt{\frac{2 \pi m}{k_B}}{\left(\Gamma (-a) \rho^V \sqrt{T^V} - \rho_{sat}|_{T_i^L} \sqrt{T_i^L}\right)}^{-1}
\end{equation}

\begin{equation} \label{eq:alpha_app_Schrage}
\alpha_{app}^{Schrage}= \frac{2 \alpha^{H-K}}{\alpha^{H-K} + 2} 
\end{equation}

\noindent where $\rho^V$ and $T^V$ are calculated by averaging the values in the gas phase. $T_i^L$ is calculated at the exact point of the intersection of liquid phase and interfacial region (see Fig.~\ref{fig:domain}c). Drift velocity ($w_0$) is evaluated by averaging the $x$-component of the vapor atoms in a large bin placed in the bulk gas phase. Mass flux ($\dot{m}^{''}$) is calculated by multiplying the drift velocity with the gas density evaluated at the same bin. Details of the calculation of uncertainty associated with MAC values are provided in Supplementary Material.

\section{Results and Discussion}

\subsection{Equilibrium simulations}
\label{sec:eqm}

\begin{figure} [h!]
\includegraphics[width=3.2in]{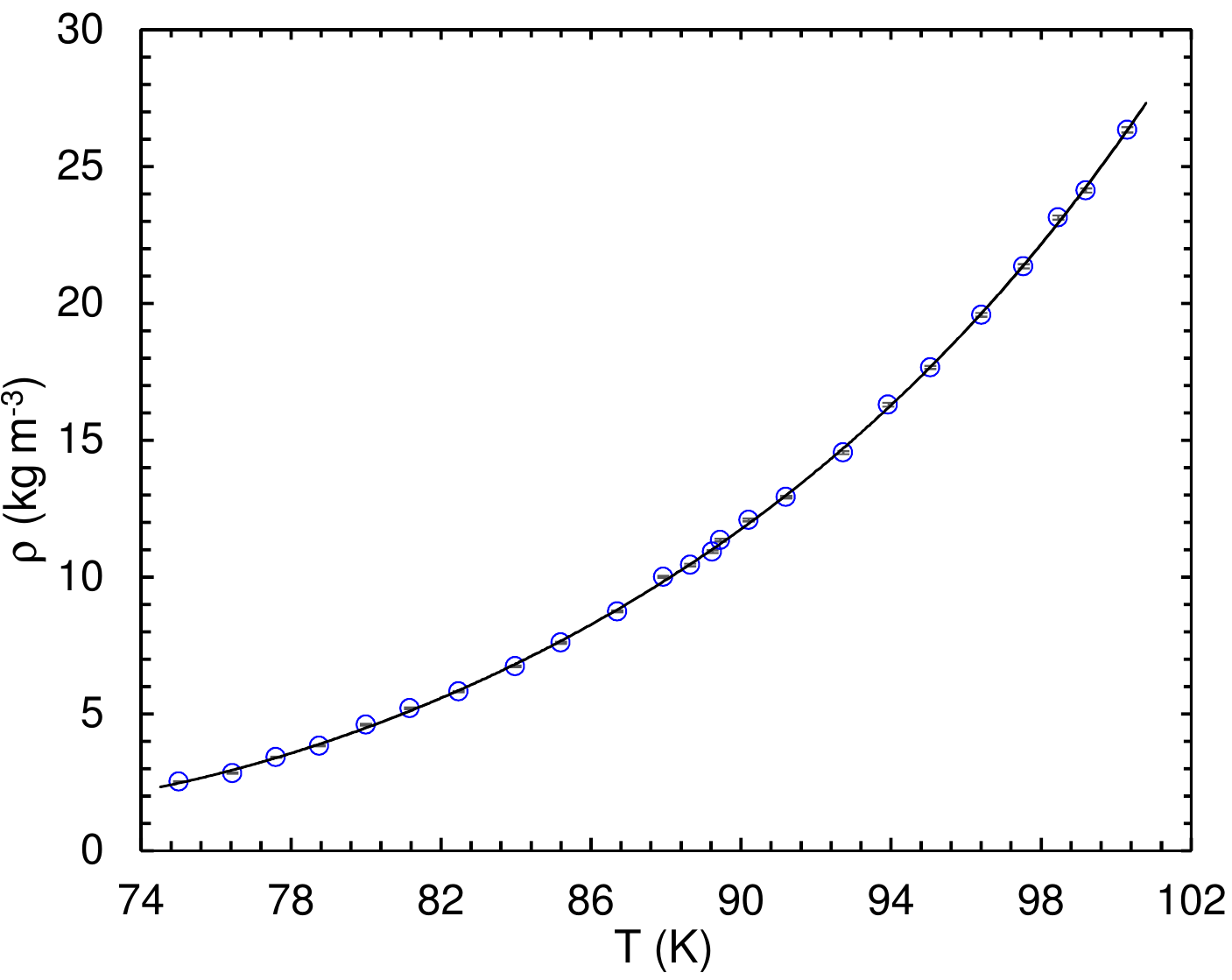}
\centering
\caption{\label{fig:eqm_rho_vap} Saturated vapor density of argon as a function of temperature. Blue circles are data points and the uncertainty of the data is smaller than the size of the circles. Solid line is the fourth order polynomial fit to the data. Addition of further data points has negligible effect on the polynomial fit.}
\end{figure}

At the nanoscale, interfacial and surface forces start to dominate over body forces, which leads to the severe deviations from continuum predictions. Near the close vicinity of the walls, molecular layering (density fluctuations) within the liquid is observed \cite{heslot1989}. Fluid properties such as density and viscosity \cite{vo2015} deviate from their bulk fluid properties. Moreover, free molecular flow or transition regime may be present in vapor phase if the mean free path of a molecule/atom is comparable to the size of the nano-conduit. Considering all these factors, equilibrium properties of a fluid in a nanoscale system can be dependent on the system itself. Therefore, saturated fluid properties of argon in our computational setup are determined from our equilibrium simulations similar to the approaches in previous studies \cite{holyst2015,liang2017}. Equilibrium (saturation) vapor density could be estimated from a fourth order polynomial fit to the saturation data but the fit coefficients depend on the number and location of data points used. Uniformly distributed data points between 80$\unit{K}$ and 100$\unit{K}$ were added one by one until the there was no further change in the corresponding fit coefficients. We observed that data at the lower end of the temperature range have the highest potential to change the fit coefficients hence it was critical to include additional data points below 80$\unit{K}$. Figure~\ref{fig:eqm_rho_vap} shows the density values of vapor obtained from 24 equilibrium simulations conducted at different temperatures.

\subsection{Near-equilibrium simulations}
\label{sec:near-eq}

\begin{figure} [h!]
\noindent
\makebox[\textwidth]{\includegraphics[width=6.5in]{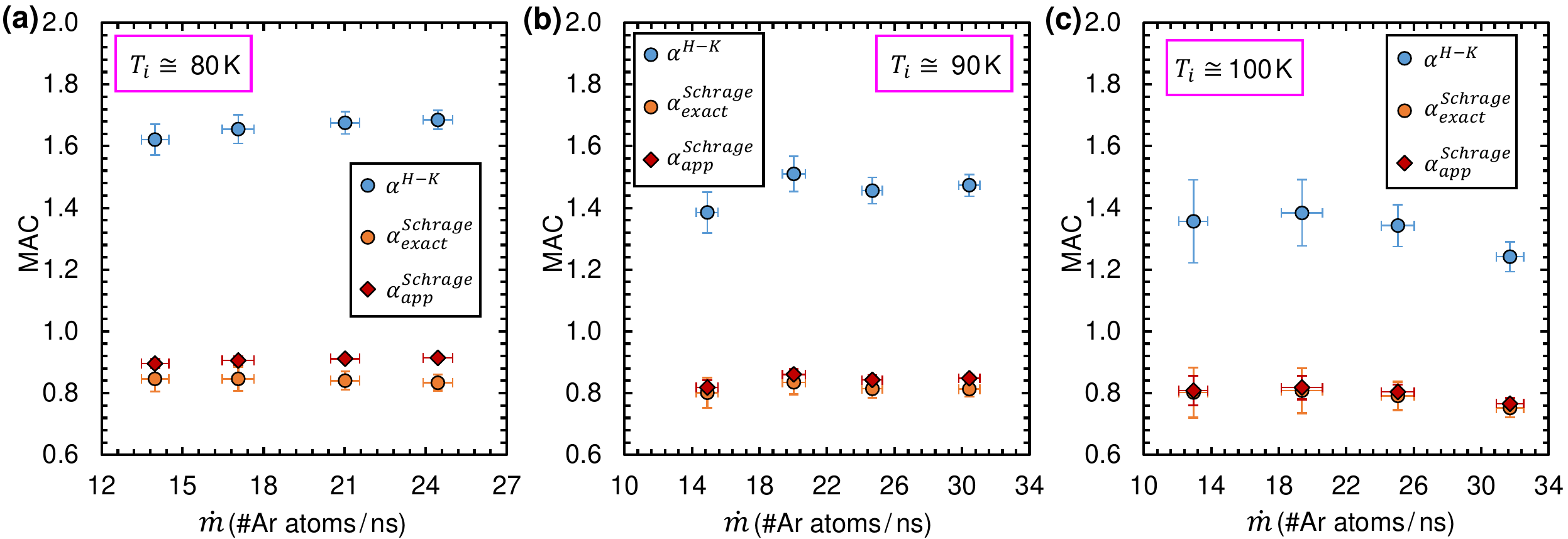}}
\caption{\label{fig:MAC_near_eq} Mass accommodation coefficients vs. condensation rate during near-equilibrium (small condensation rate) simulations.}
\end{figure}

Using Eqs.~\eqref{eq:alpha_HK}--\eqref{eq:alpha_app_Schrage}, MAC values are determined for three systems equilibrated at the temperatures of 80$\unit{K}$, 90$\unit{K}$, and 100$\unit{K}$. These are then subjected to small cooling/heating rates ($\dot{q}=0.2-0.6\unit{nW}$) during near-equilibrium simulations. Resultant MAC values on the condensing interface are reported in Fig.~\ref{fig:MAC_near_eq} as a function of the condensation rate.

\noindent During near-equilibrium condensation simulations, Schrage's exact and approximate relations yielded MAC values around 0.8-0.9. H-K equation, on the other hand, yielded MAC values between 1.3-1.7. A value higher than unity appears to contradict physical definitions (\textbf{Definitions 2-5} discussed earlier) of MAC. The most commonly used \textbf{Definition 3} suggests that if MAC is higher than 1, the condensed flux is greater than the incident flux which is a violation of mass conservation. MAC $>$ 1 may be attributed to: (i) the assumed equality of coefficients; a simplification made on the original H-K equation (Eq.~\eqref{eq:HK}), or (ii) neglect of drift velocity  or (iii) an incorrect definition of MAC. However, values exceeding unity were reported even if the evaporation and condensation coefficients were not assumed to be equal \cite{persad2016}. Drift velocity correction proposed by Schrage always yields MAC values smaller than unity in our simulations. MAC values computed from both exact and approximate Schrage relations in the current study are in good agreement with the values reported in the literature \cite{yasuoka1994argon,tsuruta1999,anisimov1999,rosjorde2001,nagayama2003}.

\begin{figure}[h]
\noindent
\makebox[\textwidth]{\includegraphics[width=6.5in]{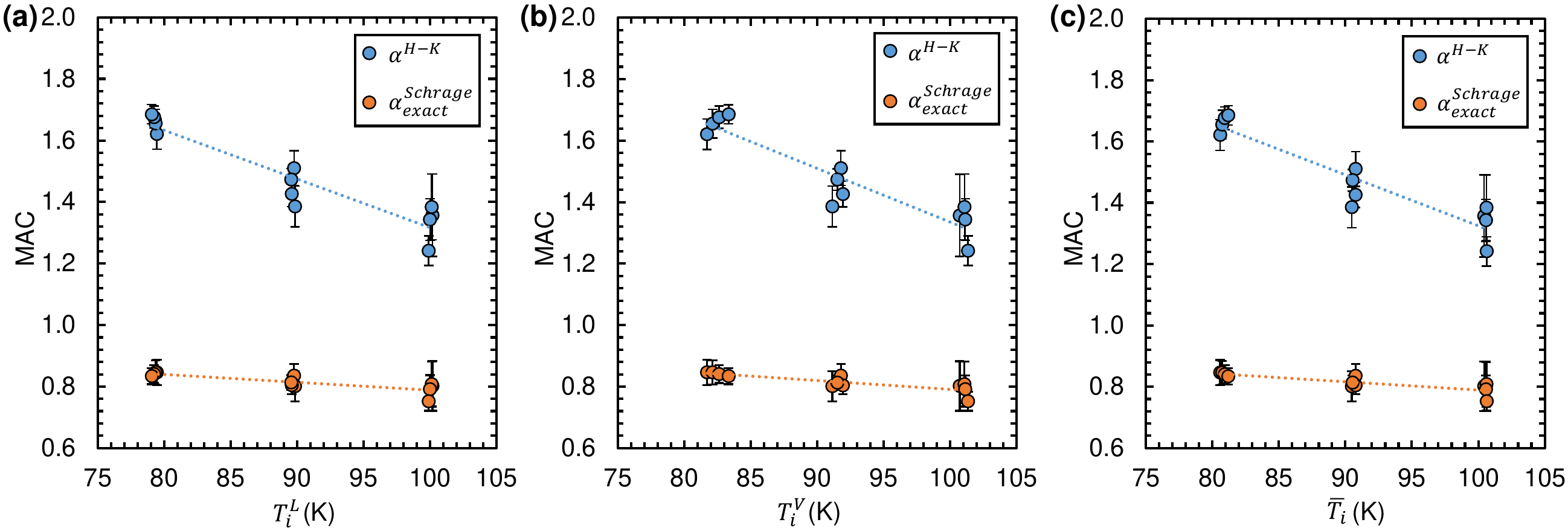}}
\caption{\label{fig:MAC_at_dif_T} Mass accommodation coefficients as the functions of (a) interfacial liquid temperature, (b) interfacial vapor temperature, and (c) average interface temperature. Dot-lines show the linear fit to the data.}
\end{figure}

Two trends are evident from Fig.~\ref{fig:MAC_near_eq}. First, MAC decreases with increasing temperature. This observation is well-known and reported many times in the literature \cite{matsumoto1998,anisimov1999,nagayama2003,ishiyama2004,tsuruta2004,tsuruta2005,holyst2009,cheng2011,nagayama2015,liang2017}. Second, the coefficients calculated from approximate and exact Schrage relations converge at high temperatures. This is due to a systemic variation in $a$, the ratio of drift velocity to the mean thermal velocity. The approximate Schrage relation is built on the assumption that the drift velocity is small compared to the mean thermal velocity. With increasing equilibration temperature, vapor velocity generally decreases due to an increase in density. This effectively results in a reduction in $a$ with temperature. Hence, the coefficients calculated from the approximate Schrage relation are nearly identical to the values calculated from the exact Schrage relation at high temperatures. 

Figure~\ref{fig:MAC_at_dif_T} demonstrates MAC values as functions of interfacial liquid, interfacial vapor, and average interface temperatures. MAC values predicted by the current study decrease with increasing temperature as expected. There is a strong temperature dependence in MAC calculated from the H-K equation, while a weak dependence is observed for MAC values calculated from the exact Schrage equation. MAC values calculated from the approximate Schrage equation have an intermediate dependence and they are not shown in Fig.~\ref{fig:MAC_at_dif_T} for brevity.

\subsection{Non-equilibrium simulations}
\label{sec:non-eq}

\begin{figure} [t!]
\center
\includegraphics[width=5.5in]{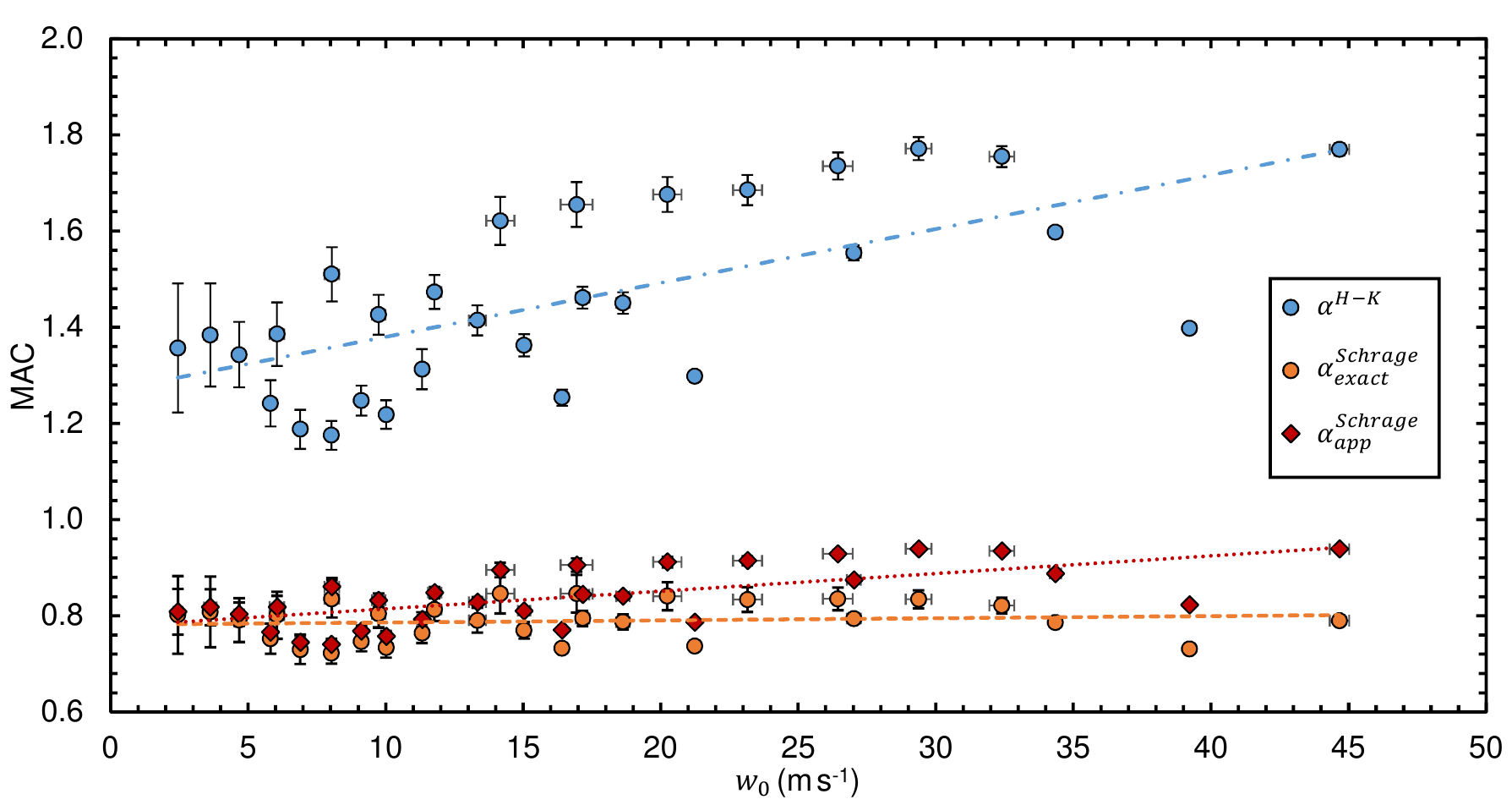}
\caption{\label{fig:rate_dep} Mass accommodation coefficients as a function of drift velocity during both near-equilibrium and non-equilibrium simulations. Dash, dot, and dash-dot lines are the linear fits to the data of MACs calculated based on exact Schrage relation, approximate Schrage relation and H-K equation, respectively. The data points are not marked for the different temperatures to avoid confusion.}
\end{figure}

\begin{figure} [b!]
\center
\includegraphics[width=4.8in]{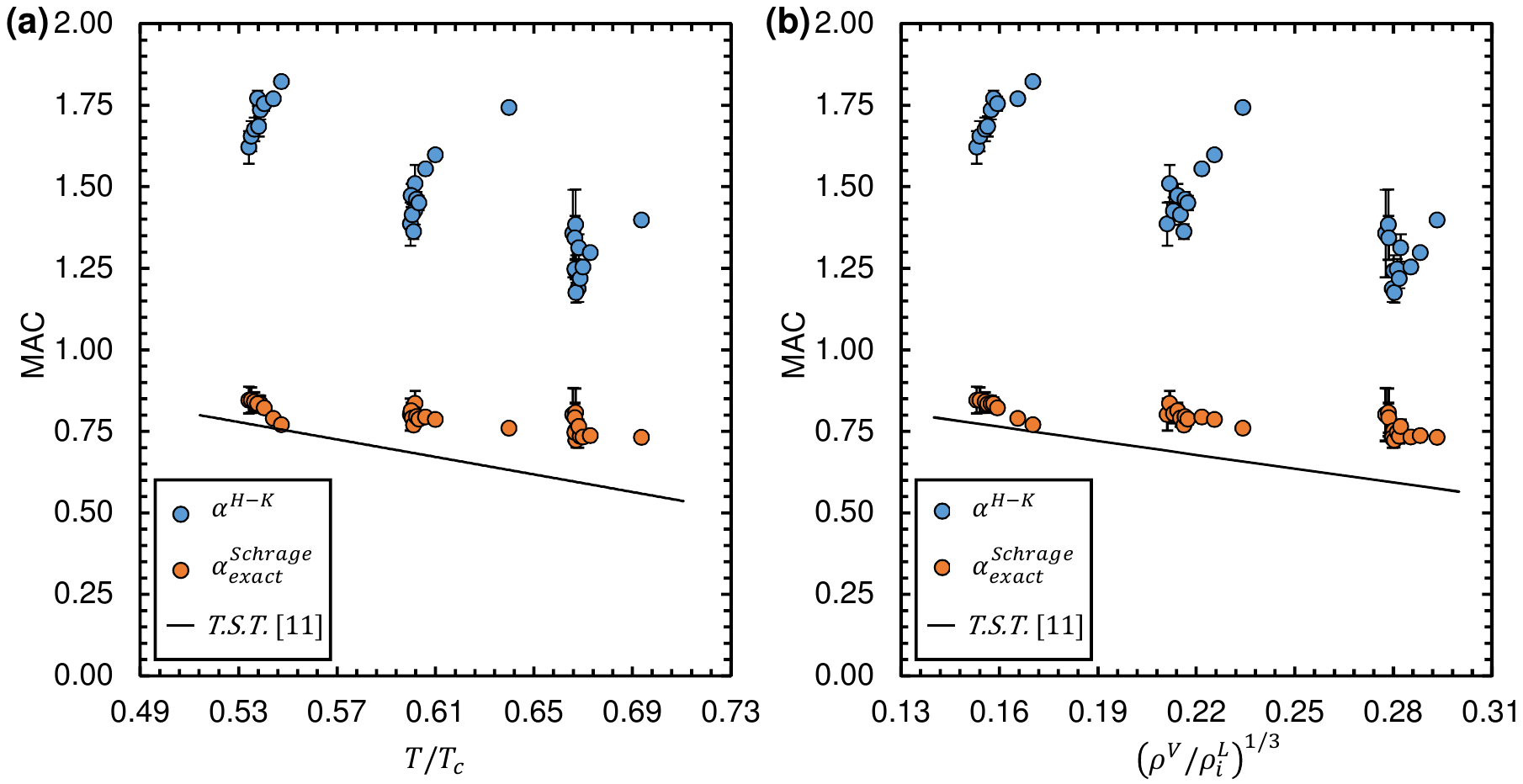}
\caption{\label{fig:TST} Mass accommodation coefficients as the functions of (a) reduced temperature, and (b) translational length ratio. Solid lines show the predictions of transition state theory (T.S.T). Critical temperature ($T_c$) of argon is $150.86 \unit{K}$.}
\end{figure}

Although the derivation of kinetic models is dependent on a near-equilibrium assumption, these equations are still used by researchers and engineers to estimate the phase change rate in various applications. Therefore, we determined the MAC values even for the cases while phase change rate was not low. After the equilibration period, non-equilibrium simulations are conducted with the cooling (and heating) rates of $\dot{q}=0.7-2.0\unit{nW}$. Surprisingly, we find that some of the MAC values vary with the heating/cooling rate which is directly proportional to both the phase change rate and drift velocity. In order to understand this behavior, MAC values are plotted as a function of drift velocity in Fig.~\ref{fig:rate_dep}. Linear fits to the scattered data reveal that MAC values calculated from H-K equation and approximate Schrage relation increase with increasing drift velocity of the gas phase, whilst exact Schrage relation yields MAC values which do not exhibit considerable variation. Therefore, exact Schrage relation is able to adjust itself during a strong phase change process \textit{via} the drift velocity correction factor.

Tsurata \textit{et al.} \cite{nagayama2003,tsuruta2005} proposed a general expression for the condensation coefficient based on transition state theory. Comparison of MAC values of the current study with the ones obtained from the expression of Tsurata \textit{et al.} is reported as the functions of both reduced temperature, $T/T_c$, and translational length ratio, $(\rho^V/\rho_i^L)^{1/3}$, in Fig.~\ref{fig:TST}. Mass accommodation coefficients calculated from exact Schrage relation is closest to the theoretical prediction. Coefficients calculated based on approximate Schrage relation is similar or slightly higher than the ones based on exact Schrage relation and they are not shown in Fig.~\ref{fig:TST} for brevity. Coefficients calculated based on H-K equation, on the other hand, almost double the predictions of transition state theory.

\subsection{Deviation between bulk and interfacial vapor properties}

\begin{figure}[h]
\noindent
\makebox[\textwidth]{\includegraphics[width=5in]{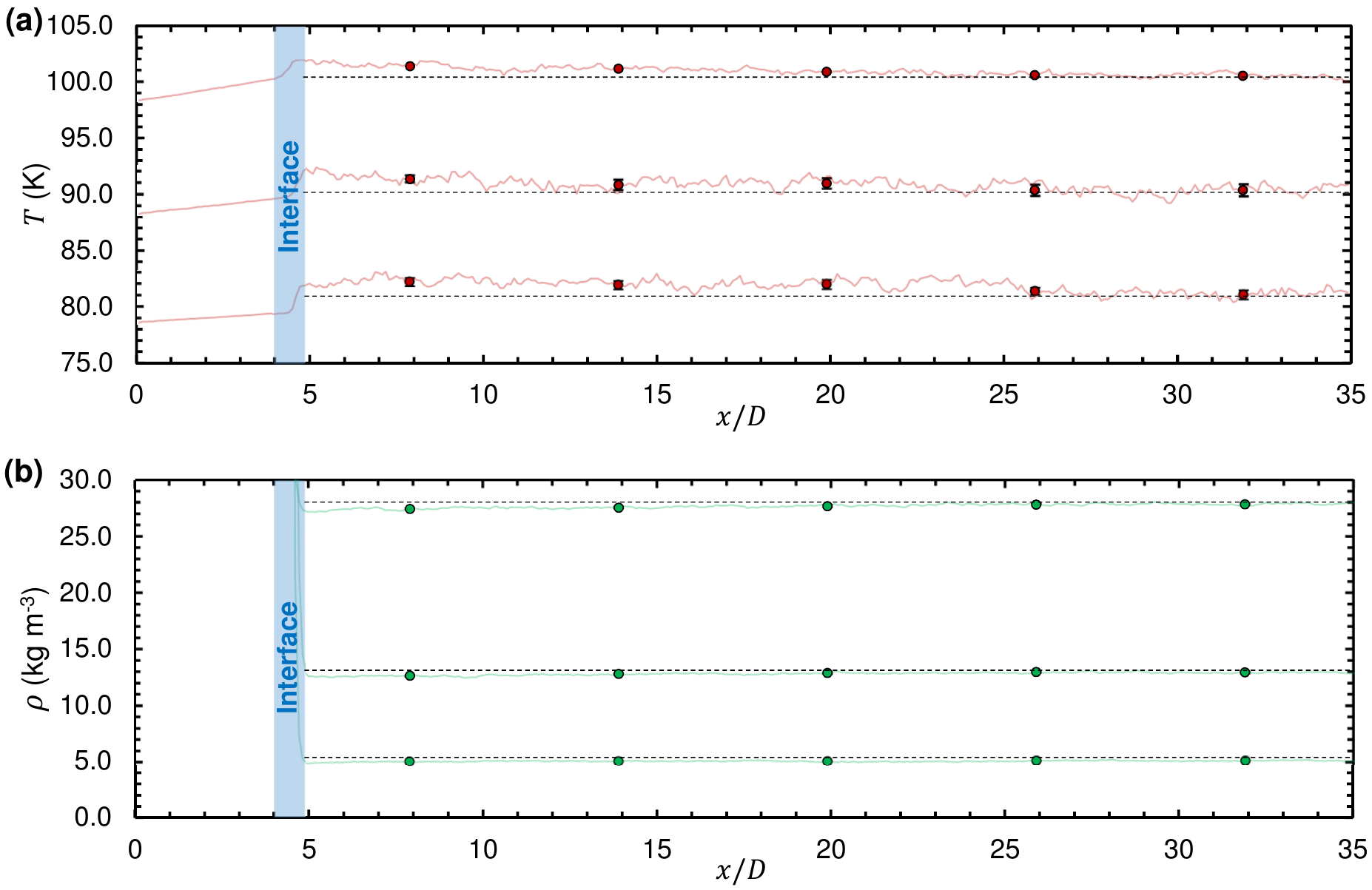}}
\caption{\label{fig:near_int_T_rho} Distributions of (a) temperature and (b) density of the gas phase near the interface for the cooling/heating rate of $0.5 \unit{nW}$. Semi-transparent solid lines show all the data. Circles show the data averaged within the intervals of $11.76 \unit{nm}$, which is nearly equal to the mean free path of an argon atom in the vapor phase around $90 \unit{K}$.}
\end{figure}

After the first evidence of an inverted temperature gradient \cite{pao1971} and temperature rise \cite{ward2001} at the condensing interface, several MD studies \cite{frezzotti2003,tsuruta2005} have been able to replicate the result. In our simulations, a temperature rise near the condensing interface and a corresponding density decrease \cite{frezzotti2003} were observed as shown in Fig.~\ref{fig:near_int_T_rho}. The temperature rise has been attributed to the latent heat of phase change \cite{ytrehus1997}. As mentioned earlier, density and temperature of the gas phase utilized in Eqs.~\eqref{eq:alpha_HK} and\eqref{eq:alpha_exact_Schrage} are determined by averaging the properties throughout bulk vapor phase. In this section, we relax the assumption of uniform vapor phase properties and retain the interfacial values. Replacing $T_V$ and $\rho_V$ with $T_i^V$ and $\rho_i^V$ in Eqs.~\eqref{eq:alpha_HK} and \eqref{eq:alpha_exact_Schrage} enables an investigation of the MAC dependence on the non-uniformity in vapor phase properties. However, temperature of the gas phase is inevitably subjected to fluctuations in the gas phase, which prevents us to determine an exact value at an exact position. Therefore, we averaged the data within enlarged bins (approximately equal to the mean free path of the gas) to eliminate the fluctuations (Fig.~\ref{fig:near_int_T_rho}). The data averaged in the bin nearest to the interface is used as the corresponding interfacial property ($T_i^V$, $\rho_i^V$). In other words, the interfacial vapor properties are averaged values from within one mean free path in the vapor phase. 

\begin{figure}[h]
\center
\includegraphics[width=5.5in]{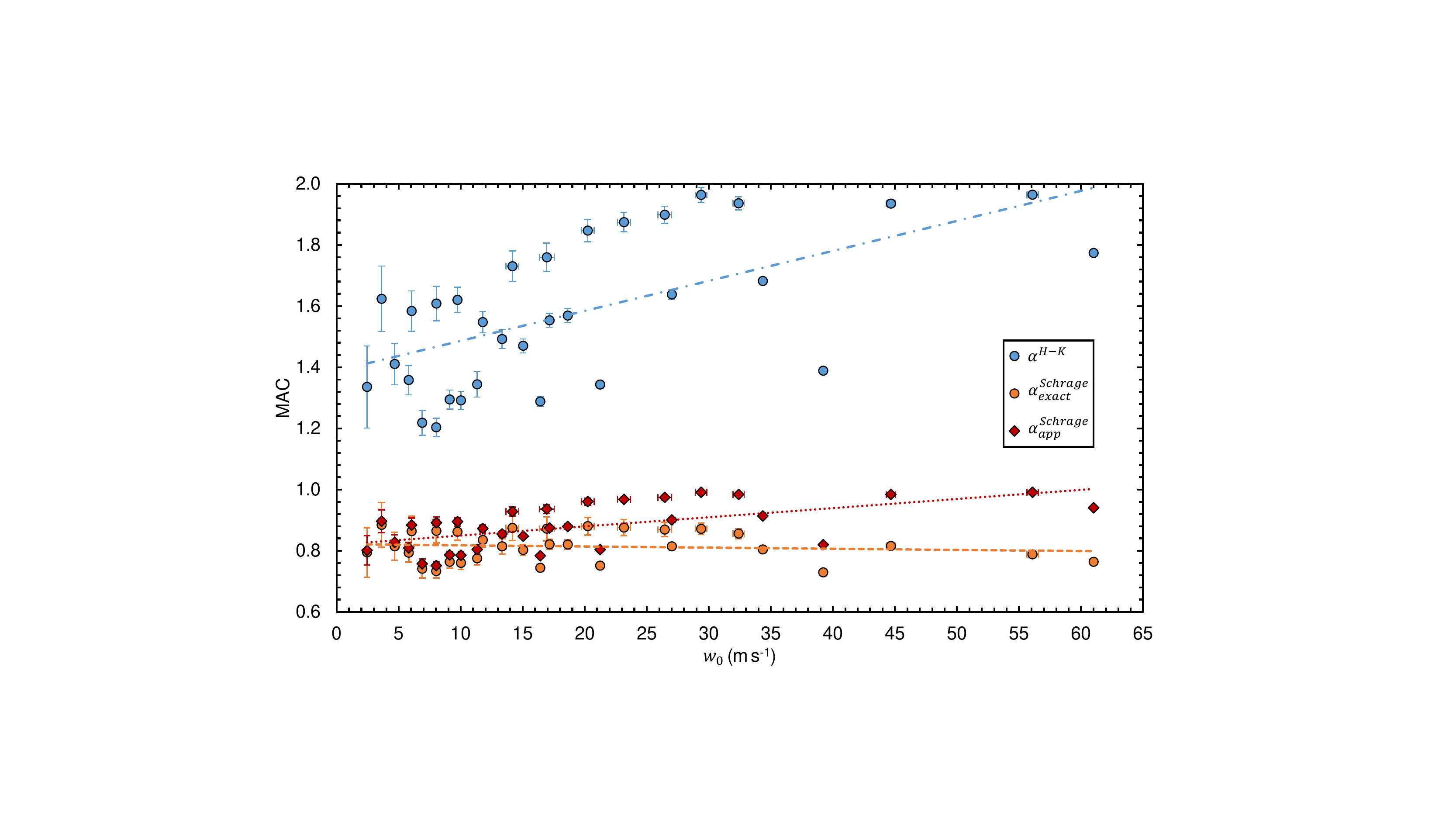}
\caption{\label{fig:rate_dep_near_vap} Mass accommodation coefficients as a function of drift velocity during both near-equilibrium and non-equilibrium simulations, when the properties of gas near the interface are utilized. Dash, dot, and dash-dot lines are the linear fits to the data of MACs calculated based on exact Schrage relation, approximate Schrage relation and H-K equation, respectively.}
\end{figure}

The resultant MAC values calculated based on the interfacial properties are shown in Fig.~\ref{fig:rate_dep_near_vap}. The trend in Fig.~\ref{fig:rate_dep_near_vap} (computed from interfacial properties) appears similar to Fig.~\ref{fig:rate_dep} (computed from bulk values) but with one important distinction: $\alpha_{exact}^{Schrage}$ no longer a constant but reduces with drift velocity. This suggests that while the exact Schrage equation is able to account only a part of the deviation from equilibrium through its macroscopic drift correction factor. It cannot effectively account for a systemic variation in interfacial properties from the bulk values. 

A monotonic increase in the temperature rise is observed with an increasing heating rate. A similar monotonic decrease is observed in the case of density. Hence, the deviation from equilibrium is portrayed by two separate effects: (i) an increase in macroscopic drift velocity and (ii) a nanoscopic variation in vapor properties near the interface. The exact Schrage expression (Eq.~\eqref{eq:alpha_exact_Schrage}) does not account for the latter and an additional correction may be necessary. In comparison, the Hertz-Knudsen equation (Eq.~\eqref{eq:alpha_HK}) does not account for either of the non-equilibrium effects.

In general, MAC values have a tendency to increase when the properties of gas near the interface are used instead of its bulk properties. The reason of this behavior is purely mathematical and can be understood when Eqs.~\eqref{eq:alpha_HK} and \eqref{eq:alpha_exact_Schrage} are examined. The temperature rise near the interface decreases the value of MAC while the reduction in density has the opposite effect. Due to the non-linear dependence on temperature in Eqs.~\eqref{eq:alpha_HK} and \eqref{eq:alpha_exact_Schrage} the effect of density reduction is greater than the effect of temperature rise and there is an overall increase in MAC. For the 80$\unit{K}$ cases, the MAC variation due to density reduction effect is almost 4 times greater than the variation due to temperature rise. In an alternative point of view, since these terms actually represent the condensation rate of vapor ($j^V$) as explained in Section~\ref{sec:kinetic}, condensation probability is expected rise if the condensation rate ($j^V$) increases with respect to the evaporation rate ($j^L$).

\begin{figure}[h]
\center
\includegraphics[width=3.2in]{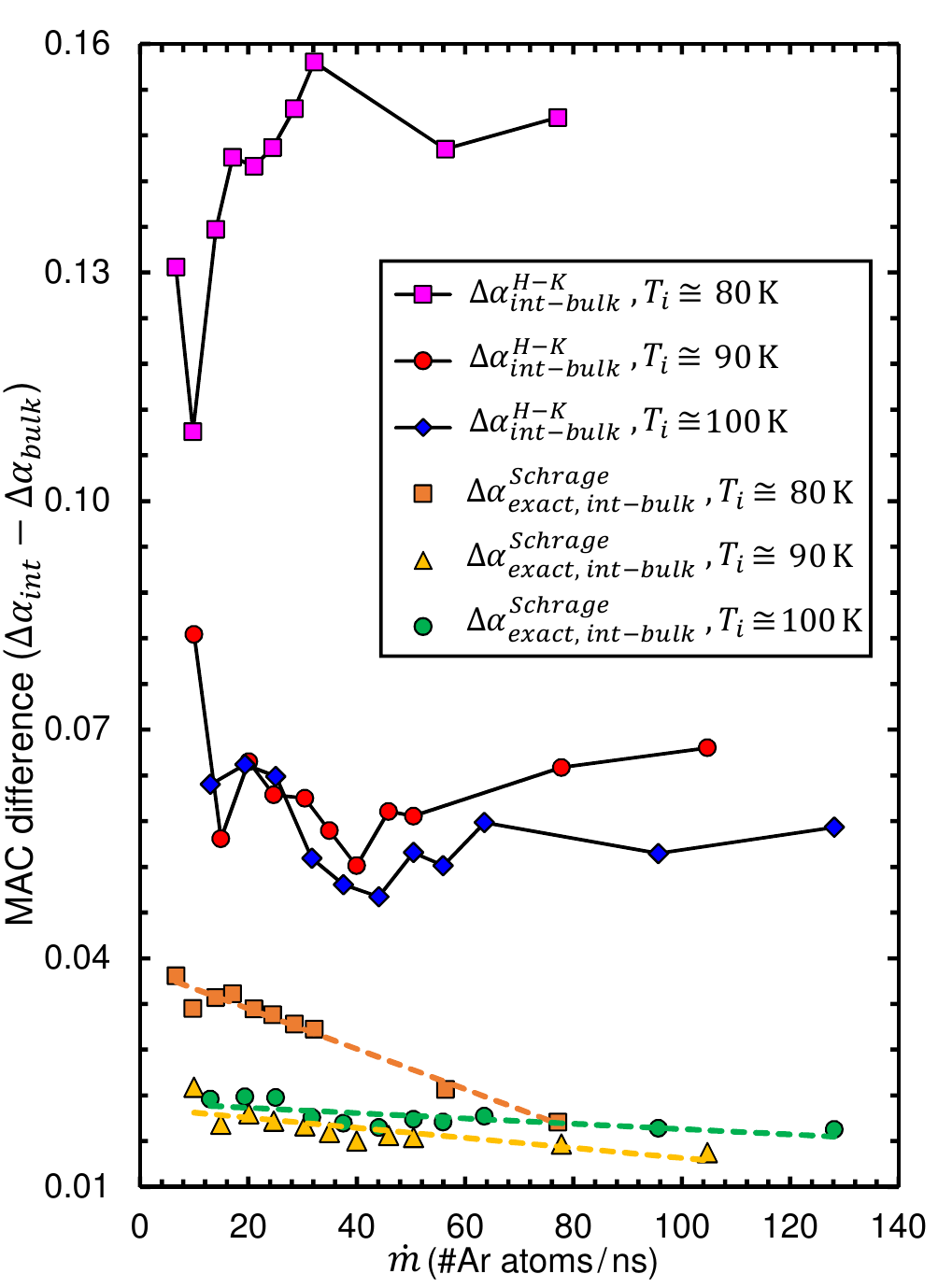}
\caption{\label{fig:delta_MAC} The difference between MAC values computed using interfacial properties ($\Delta\alpha_{int}$) and bulk properties ($\Delta\alpha_{bulk}$) as a function of phase change (condensation) rates at different temperatures. Solid lines connecting data points of $\Delta\alpha_{int-bulk}^{H-K}$ are inserted to guide the eye. Dashed lines show the linear fits to the data points of $\Delta\alpha_{exact, int-bulk}^{Schrage}$. The slopes of the linear fits reduce with increasing saturation temperatures}
\end{figure}

To characterize the effect of interfacial properties, the temperature rise and density reduction are computed for the different temperature and phase change rate conditions. To further reduce the error from random fluctuations, linear fits to the monotonically varying data are used to extract the interface vapor temperature ($T_i^V$) and interface vapor density ($\rho_i^V$) based on the previously computed bulk values. The difference between MAC values computed using interfacial properties and bulk properties ($\Delta\alpha_{int-bulk}$) are shown in Fig.~\ref{fig:delta_MAC}. $\Delta\alpha_{int-bulk}^{H-K}$ is generally greater than $\Delta\alpha_{exact,int-bulk}^{Schrage}$ but does not show a systemic variation with phase change rate. On the other hand, $\Delta\alpha_{exact,int-bulk}^{Schrage}$ shows a clear decrease with increasing phase change rate. The increased vapor motion at higher phase change rates reduces the non-uniformity in vapor properties leading to a reduced $\Delta\alpha_{exact,int-bulk}^{Schrage}$. 
Linear fits to the reducing trend are shown by the dotted lines in  Fig.~\ref{fig:delta_MAC}. Interestingly, the slope of the lines reduce with saturation temperature. This suggests that $\Delta\alpha_{exact,int-bulk}^{Schrage}$ is dependent on both phase change rate and saturation temperature.
At low nominal phase change rates or low saturation temperatures, the deviation in $\alpha_{exact}^{Schrage}$ attributable to the assumption of uniform vapor properties is non-negligible and can be as high as 0.04. However, at high phase change rates and/or high saturation temperatures, it may not be worthwhile to differentiate between interfacial and bulk vapor properties. 

\section{Summary and Conclusion}
Kinetic theory of phase change is reviewed from a fundamental standpoint and corresponding assumptions built into the most commonly used expressions for liquid-vapor phase change, Hertz-Knudsen (H-K) and Schrage relations, are discussed. Additional attention is provided to the departure from equilibrium, which alters the macroscopic vapor (drift) velocity and the thermophysical properties in the vapor. The much disputed and controversial evaporation and condensation coefficients, commonly assumed equal and termed the mass accommodation coefficient (MAC) are discussed and inconsistencies in prior definitions are highlighted.

Using a steady state molecular dynamics simulation of a phase change driven nanopump originally developed by \cite{akkus2019molecular}, the condensation process at a flat liquid vapor interface is investigated. Instead of setting a definition for MAC \emph{a priori}, the value is calculated directly from the kinetic theory relationships. The simulation data is used to calculate an explicit value of MAC from three different kinetic theory expressions: H-K equation ($\alpha^{H-K}$), approximate Schrage equation ($\alpha_{app}^{Schrage}$) and the exact Schrage equation ($\alpha_{exact}^{Schrage}$). Simulations are conducted for a range of heating/cooling rates from 0.2 - 2.0 nW at temperatures with the system saturated at 80$\unit{K}$, 90$\unit{K}$, and 100$\unit{K}$. The results are summarized below:

\begin{enumerate}

\item $\alpha^{H-K}$ is generally above unity for all cases tested. Values greater than unity violate conservation laws with respect to most common physical definitions outside the kinetic theory framework. 

\item MAC values from the Schrage expressions ($\alpha_{app}^{Schrage}$ and $\alpha_{exact}^{Schrage}$ ) are between 0.8 and 0.9. The difference between $\alpha_{app}^{Schrage}$ and the $\alpha_{exact}^{Schrage}$ expressions collapses as the saturation temperature is increased from 80$\unit{K}$ to 100$\unit{K}$.

\item MAC values from all expressions ($\alpha^{H-K}$, $\alpha_{app}^{Schrage}$ and $\alpha_{exact}^{Schrage}$) decrease with saturation pressure but the temperature dependency of $\alpha^{H-K}$ is nearly 6 times greater than the Schrage expressions.

\item $\alpha_{exact}^{Schrage}$ values are the closest match to the predictions from Transition State Theory \cite{nagayama2003,tsuruta2005}.

\item The deviation from equilibrium is characterized by two separate effects: (i) an increase in macroscopic drift velocity and (ii) a non-uniformity in vapor properties near the interface.

\item If the vapor properties are assumed uniform and the bulk vapor properties are used to determine MAC, there is no noticeable change in the $\alpha_{exact}^{Schrage}$. While both $\alpha_{app}^{Schrage}$ and $\alpha^{H-K}$ increase with drift velocity, the H-K based values have a greater sensitivity.

\item When the interfacical vapor properties are used to determine MAC, there is a general increase in the value of MAC; $\Delta\alpha_{int-bulk} = \alpha_{int} - \alpha_{bulk}>0$. $\Delta\alpha_{int-bulk}^{H-K}$ is greater than $\Delta\alpha_{exact,int-bulk}^{Schrage}$ but does not vary with the phase change rate. On the other hand, $\Delta\alpha_{exact,int-bulk}^{Schrage}$ decreases with phase change rate. The rate of decrease is dependent on the saturation temperature.

\item The Hertz-Knudsen equation does not account for either the macroscopic drift velocity or the nanoscopic variation in vapor properties near the interface. However, the exact Schrage equation accounts for the macroscopic drift velocity but not the nanoscopic variation in vapor properties near the interface. 

\end{enumerate}

In this study, MAC is computed directly from the kinetic theory expressions. As the name suggests, MAC is simply a coefficient that was originally introduced to make the equations match experimental data. In our case, we use the coefficient to match the equations with numerical simulation data. We urge that in the future when MAC values are discussed, the researcher also details the corresponding kinetic theory model along with the simplifications or explicitly state the physical definition used to characterize the coefficient without the aid of kinetic theory. If the latter method is used, it is important to note that the MAC values are then limited to just that particular definition and cannot be interchangeably used with MAC values from alternative definitions. 

\section*{Supplementary Material}

See the supplementary material for the details of molecular dynamics simulations and uncertainty analysis of the data.

\section*{Acknowledgments}

None.

\section*{Data Availability}

The data that support the findings of this study are available from the corresponding author upon reasonable request.

\addcontentsline{toc}{section}{References}

\bibliographystyle{unsrt}

\bibliography{references}

\addcontentsline{toc}{section}{Competing_interests}


\end{document}

%% file: definitions.tex
\definecolor{darkblue}{rgb}{0,0,0.8}
\definecolor{darkgreen}{rgb}{0,0.5,0}

\long\def\symbolfootnote[#1]#2{\begingroup \def\thefootnote{\fnsymbol{footnote}}\footnote[#1]{#2} \endgroup} 

\renewcommand{\exp}[1]{\text{$\hspace{0.0cm}$exp$\,#1$}}

